\documentclass[preprint,aps]{revtex4}
\usepackage{graphicx}

\begin{document}

\title{Correlation energy of a homogeneous dipolar Fermi gas}
\author{Bo Liu, Lan Yin}
\email{yinlan@pku.edu.cn} \affiliation{School of Physics, Peking
University, Beijing 100871, China}
\date{\today}

\begin{abstract}
We study the normal state of a 3-$d$ homogeneous dipolar Fermi gas
beyond the Hartree-Fock approximation.  The correlation energy is
found of the same order as the Fock energy, unusually strong for a
Fermi-liquid system.  As a result, the critical density of
mechanical collapse is smaller than that estimated in the Hartree-Fock
approximation.  With the correlation energy included, a new energy functional
is proposed for the trapped system, and its property is explored.
\end{abstract}
\pacs{03.75.Ss, 05.30.Fk, 31.15.es, 67.85.Lm}

\maketitle

\section{Introduction}
The successful creation of KRb polar molecules has provided a new
many-body system to explore \cite{Carr}.  A dipolar Fermi gas is
quite different from a typical Fermi gas or an electron gas due to the long-range
and anisotropic interaction between the dipoles.  Theoretical studies on
various properties of the dipolar Fermi gas have been carried out
\cite{Baranov}.  The Fermi surface is deformed into an elliptical shape
\cite{Pu} in the normal state of a homogeneous system.  A $p$-wave
superfluid state was proposed \cite{You}.  So far most theoretical
studies were done in the Hartree-Fock (HF) approximation which underestimates
correlation effects.  In this paper, we go beyond the HF
approximation to obtain the correlation energy of a
homogenous dipolar Fermi gas in the normal state, and use it in
the density-functional description of inhomogeneous cases.

This paper is organized as follows.  In section II, the correlation
energy of a homogeneous dipolar Fermi gas is obtained in the
perturbation theory.  We find that the ground state energy is
lowered by the correlation effect, leading to a significantly smaller
critical density of mechanical collapse than the HF result.  In
section III, the expression of the correlation energy
is used to construct the energy functional of a trapped
dipolar Fermi gas in the local density approximation.  This energy
functional is shown to satisfy a virial theorem. The critical total number of polar
molecules for mechanical collapse in the trap is obtained.  Conclusions
are given in section IV.

\section{A homogeneous dipolar Fermi gas}

We consider a homogeneous dipolar Fermi gas with mass $m$ and
permanent electric dipole moment $d$ polarized along the $z$-axis.
The uniform system is described by the Hamiltonian
\begin{equation}
H=\sum_{\bf k}\epsilon_k a^\dagger_{\bf k} a_{\bf k}
+{1\over 2V^2}\sum_{{\bf q}\ne 0} \sum_{{\bf k}_1{\bf
k}_2}V_{\bf q} a^{\dagger}_{{\bf k}_1+{\bf q}} a^{\dagger}_{{\bf
k}_2-{\bf q}}a_{{\bf k}_2} a_{{\bf k}_1}, \label{Hamiltonian}
\end{equation}
where $a_{\bf k}$ and $a^\dagger_{\bf k}$ are the fermion annihilation and creation
operators, $\epsilon_k=\hbar^2k^2/(2m)$ is the fermion kinetic energy,
$V_{\bf q}=4\pi d^2(3{\cos^2{\phi_{\bf q}}}-1)/3$ is the dipole-dipole
interaction for the transferred wavevector ${\bf q}$, $\phi_{\bf q}$ is the angle
between the $z$-axis and ${\bf q}$, and $V$ is the volume.

In the HF approximation, the single-particle excitation energy
and the Fermi surface are no longer isotropic due to the anisotropic
dipole-dipole interaction \cite{Pu}, and
the HF Hamiltonian is given by
\begin{equation}
H_{HF}=\sum_{\bf k}\varepsilon_{\bf k} a^\dagger_{\bf k}
a_{\bf k}-{1\over 2V^2} \sum_{{\bf k}_1,{\bf k}_2}
V_{{\bf k}_1-{\bf k}_2} n_{{\bf k}_1}n_{{\bf k}_2},
\end{equation}
where the single particle energy $\varepsilon_{\bf k}$ is given by
$$\varepsilon_{\bf k}=\epsilon_k+{1\over V}\sum_{{\bf k}'}V_{{\bf k}-{\bf k}'} n_{{\bf k}'},$$
$n_{\bf k}={\rm \theta}(\mu-\varepsilon_{\bf k})$ is the
occupation number, and the chemical potential $\mu$ is determined from the total
number constraint $N=\sum_{{\bf k}}n_{\bf k}$.  In the HF approximation,
the occupation number $n_{\bf k}$ can be solved self consistently.  To the first
order in the interaction strength, the deformed Fermi surface is given by
\cite{Shai}
\begin{equation}\label{kf}
k(\phi)=k_F+{1\over 9\pi}{md^2k_F^2\over
\hbar^2}(3\cos^2\phi-1),
\end{equation}
where $k_F$ is the Fermi wavevector of the noninteracting Fermi gas.
To the first order, this deformation does not change the volume inside the Fermi surface,
nor the total energy.  The HF approximation modifies the ground state energy starting
only in the second order \cite{Shai},
$${E_{HF}\over V}={\hbar^2k_F^5\over m}[{1\over
20\pi^2}-{1\over 405\pi^4}({m d^2k_F\over \hbar^2})^2],$$
or in terms of the Fermion density $n$,
\begin{equation}\label{HF1 Eg}
{E_{HF}\over V}={\hbar^2\over m}[{3\over 10}(6\pi^2)^{{2\over3}}n^{{5\over
3}}-{4\over 45}(6\pi^2)^{{1\over3}} ({m d^2\over \hbar^2})^2n^{7\over 3}],
\end{equation}
where $n$ is the Fermion density in real space.

Although the HF approximation is accurate for the ground state energy in the first order,
in general it is insufficient for the second or higher-order ground state energy.
Kohn and Luttinger \cite{Kohn} showed that in Fermi systems with anisotropic Fermi surfaces
due to interaction, the second-order ground state energy consists of two parts, one
from the HF approximation and the other from the Brueckner-Goldstone (BG) formula.
The BG formalism is the standard perturbation theory which was used by
Lee and Yang to obtain the second-order ground state energy of a dilute Fermi
gas \cite{Lee_Yang}.
In the BG formalism, the second-order contribution to the ground-state energy
of a dipolar Fermi gas is given by
\begin{equation}
E^{(2)}_{BG}=\sum_{m\neq0} {\langle 0\mid H_1 \mid m \rangle
\langle m \mid H_1\mid 0 \rangle \over E_0-E_m},
\label{Energy21}
\end{equation}
where $\mid 0 \rangle$ and $\mid m \rangle$ are the ground
and excited states of the noninteracting Fermi gas with eigenenergies
$E_0$ and $E_m$, and $H_1$ is the dipole-dipole interaction given by
the second r.-h.-s. term in Eq. (\ref{Hamiltonian}).
This BG energy can be further written as
\begin{eqnarray}
{ E^{(2)}_{BG} \over V}&=&{1\over 2 V^3}\sum_{{\bf k}_1,{\bf
k}_2,{\bf q}} {V_{\bf q}(V_{\bf q}-V_{{\bf k}_2-{\bf k}_1-{\bf q}})
\over \epsilon_{{\bf k}_1}+\epsilon_{{\bf k}_2}-\epsilon_{{\bf
k}_1+{\bf q}}-\epsilon_{{\bf k}_2-{\bf q}}} \nonumber \\
&&\times f_{{\bf k}_1}f_{{\bf k}_2}(1-f_{{\bf k}_1+{\bf q}})(1-f_{{\bf
k}_2-{\bf q}}),\label{Energy22}
\end{eqnarray}
where $f_{\bf k}=\theta(k_F-k)$ is the fermion occupation number in
the noninteracting ground state $\mid 0 \rangle$.
The ${\bf k}$-space summations in Eq. (\ref{Energy22}) can be rescaled into
integrals over dimensionless variables,
\begin{eqnarray}
{ E^{(2)}_{BG} \over V}&=& {md^4 k_F^7\over 32\pi^7\hbar^2}\int d^3x_1d^3x_2d^3y
{(\cos^2\phi_{\bf y}-{1 \over 3})(\cos^2\phi_{\bf y}-
\cos^2\phi_{{\bf x}_2-{\bf x}_1-{\bf y}}) \over x_1^2+x_2^2
-|{\bf x_1}+{\bf y}|^2-|{\bf x_2}-{\bf y}|^2} \nonumber \\
&&\times \theta(1-x_1) \theta(1-x_2)\theta(|{\bf x_1}+{\bf y}|-1)
\theta(|{\bf x_2}-{\bf y}|-1), \label{Deng}
\end{eqnarray}
where ${\bf x}_i={\bf k}_i/k_F$ and ${\bf y}={\bf q}/k_F$.
Eq. (\ref{Deng}) indicates that the BG energy density is proportional
to $d^4n^{7/3}$.  We obtain the coefficient by Monte Carlo integration \cite{Press},
\begin{equation}
{E^{(2)}_{BG} \over V}=-0.66{m d^4 \over \hbar^2}  n^{7\over 3}.
\label{Energy24}
\end{equation}
By definition, the correlation energy is the difference between the true
ground state energy and the ground state energy in the HF
approximation.  Therefore in the second order, the correlation energy of
a dipolar Fermi gas is given by Eq. (\ref{Energy24}).  Its negative sign
indicates that the ground state energy is lowered by the correlation effect.

\begin{figure}[t]
\begin{center}
\includegraphics[width=8cm]{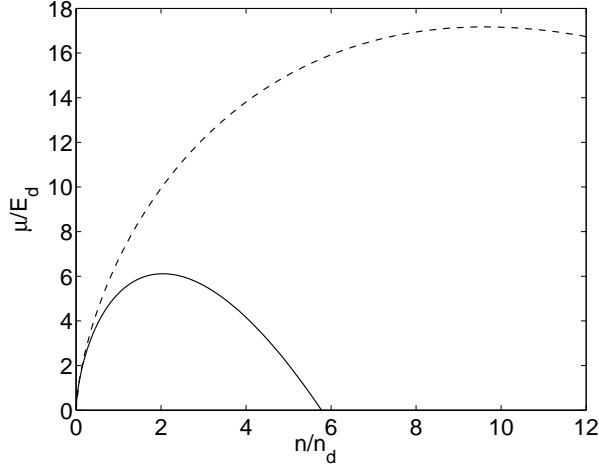}
\caption{Chemical potential $\mu$ as the function of the fermion density $n$,
where the unit of $\mu$ is $E_d\equiv\hbar^6/(m^3 d^4)$ and the unit of $n$
is $n_d\equiv[\hbar^2/ (m d^2)]^3$.  The solid and dashed lines are the chemical potentials
with and without the contribution from correlation energy respectively.} \label{figure1}
\end{center}
\end{figure}

The total ground-state energy of a homogeneous
dipolar Fermi gas up to the second order can be obtained by
adding Eq. (\ref{Energy24}) to Eq. (\ref{HF1 Eg}),
\begin{equation}
{E\over V}={\hbar^2\over m}[{3\over 10}(6\pi^2)^{{2\over3}}n^{{5\over
3}}-1.01({m d^2\over \hbar^2})^2 n^{7\over 3}].
 \label{Energy} \label{soge}
\end{equation}
Since the second r.-h.-s. term in Eq. (\ref{soge}) is negative, and its magnitude
increases faster than the kinetic energy when the density increases,
there exists a critical density $n_c$ for mechanical collapse, $(\partial \mu / \partial n)_{n_c}=0$,
where the chemical potential $\mu$ is given by
\begin{equation}
\mu={\partial E \over \partial N}={\hbar^2\over m}[{1\over 2}(6\pi^2)^{{2\over3}}
n^{{2\over 3}}-2.36({m d^2\over \hbar^2})^2n^{4\over 3}].
\label{mu}
\end{equation}
As shown in Fig. \ref{figure1}, the chemical potential reaches the maximum at
the critical density $n_c=2.04 (\hbar^2 /  m d^2)^3$,
whereas the simple HF result implies $n_c=9.97[\hbar^2/( m d^2)]^3$.
Beyond this critical density $n_c$, the compressibility turns
negative and the system is no longer stable against
mechanical collapse.  We should point out that the perturbation method is valid when the absolute value of the interaction
energy is smaller than the kinetic energy, $n<9.58 (\hbar^2/md^2)^3$
as indicated by Eq. (\ref{soge}).  At $n_c=2.04 (\hbar^2 /  m d^2)^3$, the perturbation still works,
but higher order terms may be needed to produce more quantitatively accurate results.
Nonetheless, the correlation energy effectively makes the system more vulnerable
towards mechanical collapse.

In the experiment on KRb molecules, the dipole moment for the singlet
rovibrational ground state is $d=0.566(17)$ Debye  \cite{K.K.Ni}.  Our calculation suggests
that the critical density for mechanical collapse in this case is approximately
$n_c\simeq9.23\times10^{12} cm^{-3}$.
In other references, this critical density is estimated to be $1.48\times10^{14} cm^{-3}$
in the variational approach \cite{sogo}, $4.89\times10^{12} cm^{-3}$ from the
zero-sound analysis \cite{Shai}, and $4.80\times10^{12} cm^{-3}$ in the collective-mode
study \cite{Chan}.  The highest molecule density in the current experiment is
about $0.3\times10^{12} cm^{-3}$ \cite{K.K.Ni1} when the temperature is at $1.4$ times the Fermi
temperature.  It is hopeful that these theoretical results can be tested in future experiments
with higher molecule density achieved.

\section{Trapped dipolar Fermi gas}
The Hamiltonian of a dipolar Fermi gas in a harmonic trap is given by
\begin{eqnarray}
H&=& \int d^3r \psi^{\dagger}({\bf r})[{-\hbar^2\nabla^2\over
2m}+U_{ho}]\psi({\bf r}) \nonumber \\
&+& {1\over 2} \int\int d^3 r d^3 r^{\prime}\psi^{\dagger}({\bf r})
\psi^{\dagger}({\bf r}^{\prime})V_{dd}({\bf r}-{\bf r}^{\prime})
\psi({\bf r}^{\prime})\psi({\bf r}), \nonumber \\
\label{Hamilton2}
\end{eqnarray}
where the dipole-dipole interaction $V_{dd}({\bf r})$ is given by
$V_{dd}({\bf r})=(d^2/r^3)[1-3(z^2/r^2)]$, and
$\psi({\bf r})$ is the fermion field operator.  In the following
we consider an axially symmetric trap with the trapping potential given by
$U_{ho}=m(\omega_\rho^2 x^2+\omega_\rho^2 y^2+\omega^2_z z^2)/2$
and the trap aspect ratio given by $\lambda\equiv\omega_z/\omega_\rho$.

We consider the case that the scale of the trap is much larger than
the interparticle distance, so the local density approximation (LDA) can be
applied.  The total energy $E$ can be written as
$$E=E_{kin}+E_{ho}+E_{Har}+E_{s},$$ where $E_{kin}$, $E_{ho}$, and
$E_{Har}$ are kinetic, trap, and Hartree energies.  The energy
$E_{s}$ includes both the second-order Fock and correlation
energies.  Density functionals of these energies are given by
\begin{eqnarray*}
E_{kin}&=&{3(6\pi)^{2\over 3}\hbar^2\over 10m}\int d^3r
n^{5\over3}({\bf r}),\\E_{ho}&=&\int d^3rU_{ho}({\bf r})n({\bf r}),\\
E_{Har}&=&{1\over 2}\int \int d^3 r d^3 r^{\prime}n({\bf r})
V_{dd}({\bf r}-{\bf r}^{\prime})n({\bf r}^{\prime}),\\
E_{s}&=&-1.01{m d^4\over \hbar^2}\int d^3 r n^{7\over 3}({\bf r})
\end{eqnarray*}
where $n({\bf r})$ is the fermion density with the number constraint $N=\int d^3r n({\bf r})$.
At the ground state, the fermion density $n({\bf r})$ satisfies the energy-extreme condition,
\begin{equation} \label{EEC}
{\delta E \over \delta n}-\mu {\delta N \over \delta n}=0,
\end{equation}
from which the chemical potential $\mu=\partial E/\partial N$ can be determined,
\begin{eqnarray}
\mu &=&{\hbar^2\over 2m}(6\pi^2)^{2\over 3}n^{2\over 3}({\bf r})+
{\hbar \omega_\rho\over 2}(x^2+y^2+\lambda^2 z^2) \nonumber \\
&+&d^2\int d^3 r^{\prime}{1-3\cos^2\theta_{{\bf r}-{\bf
r^{\prime}}}\over \mid {\bf r}-{\bf r^{\prime}}\mid^3}n({\bf
r^{\prime}}) \nonumber \\&-&2.36 {m d^4\over \hbar^2} n^{4\over
3}({\bf r}). \label{density}
\end{eqnarray}

In LDA, the total energy $E$ satisfies a virial
theorem which can be proved by the scaling transformation of the
density \cite{Englert1, Englert2}
\begin{equation}
n({\bf r})\rightarrow \eta^{3+\alpha}n(\eta{\bf r}),
\label{trans1}
\end{equation}
where $\eta$ and $\alpha$ are general numbers.
Under this transformation, the total density and various parts of
the total energy $E$ transform as
\begin{eqnarray*}
N&\rightarrow& \eta^{\alpha}N, \\
E_{kin}&\rightarrow& \eta^{2+{5\over 3}\alpha}E_{kin}, \\
E_{ho}&\rightarrow&\eta^{-2+\alpha}E_{ho}, \\
E_{Har}&\rightarrow&\eta^{3+2\alpha}E_{Har},\\
E_{s}&\rightarrow&\eta^{4+{7\over 3}\alpha}E_{s}.
\end{eqnarray*}
Close to $\eta=1$, to the first order in $\delta \eta=\eta-1$,
the changes in these quantities are given by
\begin{eqnarray*}
\delta N &=& \alpha\delta \eta N  , \\
\delta E_{kin}&=&(2+{5\over 3}\alpha)\delta \eta E_{kin} ,\\
\delta E_{ho}&=& (-2+\alpha)\delta \eta E_{ho}, \\
\delta E_{Har}&=& (3+2\alpha) \delta \eta E_{Har},\\
\delta E_{s}&=& (4+{7\over 3}\alpha)\delta \eta E_{c}.
\end{eqnarray*}
Therefore the change in the total energy is given by
\begin{equation}
\delta E=\delta E_{kin}+\delta E_{ho}+\delta E_{Har}+\delta E_{s},
\end{equation}
and from the energy-extreme condition Eq. (\ref{EEC}) it also
satisfies the equation
\begin{equation}
\delta E={\partial  E \over \partial N} \delta N.
\end{equation}
From the above equations, we have
\begin{eqnarray}
\alpha N {\partial E \over \partial N}&=&
(2+{5\over 3}\alpha)E_{kin}+(-2+\alpha)E_{ho} \nonumber \\
&+&(3+2\alpha)E_{Har}+(4+{7\over 3}\alpha)E_{s},
\label{trans3}
\end{eqnarray}
which leads to two independent equations
\begin{eqnarray}
2E_{kin}-2E_{ho}+3E_{Har}+4E_{s}&=&0, \label{trans4} \\
{5\over 3} E_{kin}+E_{ho}+2E_{Har}+{7 \over 3}
E_{s}&=&N{\partial E \over \partial N}.
\label{trans5}
\end{eqnarray}
Eqs. (\ref{trans4}) and (\ref{trans5}) together with identities
\begin{eqnarray}
d{\partial E \over \partial d}&=&2E_{Har}+4E_{s}, \nonumber \\
\omega_\rho {\partial E\over \partial \omega_\rho}&=&2E_{ho}, \nonumber  \\
m{\partial E\over \partial m}&=&E_{ho}-E_{kin}+E_s, \nonumber
\end{eqnarray}
yield
\begin{eqnarray}
d{\partial E\over \partial d}&=& {4 \over 3} m{\partial E\over \partial m}, \nonumber \\
\omega_\rho {\partial E \over \partial \omega_\rho}&=&E+{1 \over 3} m{\partial E \over \partial m}, \nonumber  \\
N{\partial E\over \partial N}&=&{4 \over 3}E+{1 \over 9} m{\partial E \over \partial m}, \nonumber
\end{eqnarray}
indicating the total energy $E$ of the form
\begin{equation}
E(m,\omega_\rho,d,N)=\hbar \omega_\rho N^{4\over 3}g(N^{1\over 6}C_{dd}),
\label{trans6}
\end{equation}
where $C_{dd}$ is the dimensionless dipolar interaction strength,
$C_{dd}\equiv md^2/a_0\hbar^2$, $a_0=\sqrt {\hbar /m\omega_\rho}$, and $g(x)$
is a general functional of the variable $x$.

\begin{figure}[t]
\begin{center}
\includegraphics[width=8cm]{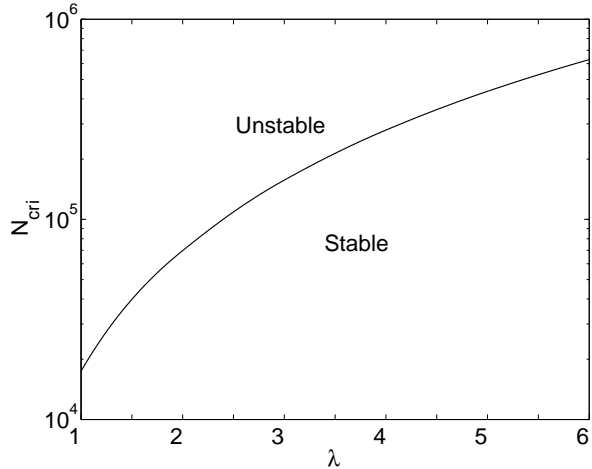}
\caption{The critical total number $N_{cri}$ of polar molecules for mechanical
collapse versus the trap aspect ratio $\lambda$ for the singlet
rovibrational ground state of KRb with fixed longitudinal trap
frequency $\omega_z/(2\pi)=109$ Hz.} \label{figure2}
\end{center}
\end{figure}

In a trap, the system also suffers mechanical collapse if
the total number of fermions exceeds a critical value.
In the experiment on KRb polar molecules \cite{K.K.Ni}, the optical
trapping frequency for the polar molecule is about $\omega_\rho/(2\pi)=109$ Hz
and the permanent electric dipole moments measured with stark spectroscopy is 0.566(17) Debye
for the singlet rovibrational ground state, corresponding to dimensionless
dipolar interaction strength $C_{dd}=0.2844$.  Our numerical results for
the experimental condition show that the critical total number of polar
molecules depends on the shape of the optical trap.  As shown in
Fig.~(\ref {figure2}), when $\omega_z$ is fixed and $\lambda$ varies, the critical total number of
polar molecules monotonically increases with the trap aspect ratio $\lambda$,
indicating that a more pronounced pancake shape of the trap can help
stabilize the dipolar Fermi gas.  Our results show that
the critical molecule number for mechanical collapse is about $1.75\times10^{4}$
for $\lambda=1$, slightly smaller than the HF result $2.15\times10^{4}$
\cite{Zhang}, implying the correlation effect
further destabilizes a trapped dipolar Fermi gas.
Again we would like to make a remark that our result about the correlation energy
given by Eq. (\ref{Energy24}) is obtained in the second-order perturbation
theory.  Better results about the correlation energy are needed to estimate
the critical molecule number for mechanical collapse with more accuracy.

\section{Conclusion}
We studied the normal state of a dipolar Fermi gas beyond the
HF approximation.  For the homogeneous system, we obtained
the ground-state energy including the correlation energy up to
second order in the dipole interaction strength, which leads to a
critical density for mechanical collapse smaller than that estimated in the
HF approximation.  For the trapped dipolar Fermi gas, we
proposed a energy functional to describe this system based on LDA.
We show that this energy functional
satisfies a virial theorem.  The instability of this system is also
investigated.  In both cases, the correlation energy makes the system
more vulnerable toward mechanical collapse.

\section*{Acknowledgment}
We would like to thank T.-L. Ho, S. Yi, and H. Zhai for helpful
discussions and other colleagues for constructive feedbacks.
This work is supported by NSFC under Grant No. 10974004,
and by Chinese MOST under grant number 2006CB921402.

\end{document}